# The optical emission nebulae in the vicinity of WR 48 (Θ Mus); True Wolf-Rayet ejecta or unconnected supernova remnant?

M. Stupar,[1,2] Q.A. Parker,[1,2] M.D. Filipović[3]

[1] *Department of Physics, Macquarie University, Sydney 2109, Australia*
[2] *Anglo-Australian Observatory, P.O. Box 296, Epping, NSW 1710, Australia*
[3] *University of Western Sydney, Locked Bag 1797, Penrith South DC, SW 1797 Australia*

**ABSTRACT**

During searches for new optical Galactic supernova remnants (SNRs) in the high resolution, high sensitivity Anglo-Australian Observatory/United Kingdom Schmidt Telescope (AAO/UKST) Hα survey of the southern Galactic plane, we uncovered a variety of filamentary and more diffuse, extensive nebular structures in the vicinity of Wolf-Rayet (WR) star 48 (Θ Muscae), only some of which were previously recognised. We used the double-beam spectrograph of the Mount Stromlo and Siding Spring Observatory (MSSSO) 2.3-m to obtain low and mid resolution spectra of selected new filaments and structures in this region. Despite spectral similarities between the optical spectra of WR star shells and SNRs, a careful assessment of the new spectral and morphological evidence from our deep Hα imagery suggests that the putative shell of Θ Mus is not a WR shell at all, as has been commonly accepted, but is rather part of a more complex area of large-scale overlapping nebulosities in the general field of the WR star. The emission comprises a possible new optical supernova remnant and a likely series of complex H II regions. Finally, we present the intriguing detection of apparent collimated, directly opposing, ionized outflows close to Θ Mus itself which appears unique among such stars. Although possible artifacts or a temporary phenomenon monitoring of the star is recommended.

## 1 INTRODUCTION

The optical broad-band or preferably narrow-band imaging morphologies of emission line structures can provide clues as to their underlying nature, especially if their shapes are pathological examples of their generic object types. Usually however, corroborating optical spectroscopy or other data provide the basis for firm identification. For example, for optical filaments or nebular emission to be classified as belonging to supernova remnants (SNRs), planetary nebulae (PN), H II regions, Wolf-Rayet shells or other exotica, a variety of optical emission line ratios and diagnostic plots have traditionally been used to assist the discrimination between the object classes (e.g. Sabbadin, Minello & Bianchini (1977), Cantó (1981) and Frew & Parker (2009)). For SNRs for example, whether they have confirmed radio counterparts or not, the optical spectra are expected to show typical diagnostic lines and ratios from the extant confirmed and optically detected SNRs (see Stupar, Parker & Filipović (2008) and references therein). SNR spectra are not generally photo-ionised. Instead the emission arises from strong, shock heating and collisional excitation as the expanding blast-wave and ejecta sweeps up and collides with the surrounding ISM. Usually, this is diagnosed via the presence of a very strong [S II] doublet at 6717 and 6731Å relative to Hα. Typically, a ratio of [S II]/Hα > 0.5 has often been used to classify nebular emission as likely arising from an SNR, especially if the associated optical morphology is pathological and there are other corroborating pieces of evidence such as radio structure, a central X-ray source or a pulsar. Such optical emission line ratios are often used to separate SNRs from H II regions and planetary nebulae (PN). Unfortunately, as lower surface brightness and more extreme, evolved examples of these different objects are increasingly uncovered from the next generation on narrow-band imaging surveys, e.g. Parker et al. (2006), Miszalski et al. (2008) these optical diagnostics are not always clear cut (Frew 2008; Frew & Parker 2009). The various object populations which exhibit optical emission line spectra can overlap in the so called Cantó diagram (Cantó 1981). Indeed, high ratios of [S II]/H$\alpha$ $\leq$ 1.0 have been confirmed in a few bona-fide, highly evolved PNe which are strongly interacting with the ISM, which is, in fact, the source of the shock-excitation (in highly evolved, see Pierce et al. 2004). Fortunately, there are other optical emission lines such as very strong [O II] at 3727Å, the Balmer lines, [O III] at 4959 and 5007Å, [N II] at 6548 and 6584Å and especially strong [O I] at 6300 and 6364Å, that can assist in assigning an SNR classification as [O I] is not prominent in PN or H II regions.

There is some confusion in these emission line ratio selection criteria, especially between SNRs and Wolf-Rayet nebula which can exhibit both morphological similarities in the form of optical filaments and filamentary shells and similarities in their optical spectra. Of course the confirmed detection of a WR star in the centre of such nebulae resolves any ambiguity. Usually, the WR nebulae (Miller & Chu 1993; Marston et al. 1994; Marston 1997) are in the form of a ring (or significant fraction of a ring) centred on the star and consisting mainly of stellar ejecta from the WR star jh65 and swept-up material from the interstellar medium. In some WR stars, the ratio of [S II] / H$\alpha$ emission lines is close to that typically seen for SNRs. This is understandable, as we know that Wolf-Rayet stars are evolved, extremely hot and luminous massive stars undergoing rapid mass loss which can form both shock-excited and photo-ionised nebula from the material discarded by the central star (Johnson & Hogg 1965). Fortunately, the progenitor star can always be identified if the nebula has a WR origin enabling discrimination against an SNR.

During our work (Stupar, Parker & Filipović 2008, 2007a,b) on uncovering new Galactic SNRs from the AAO/UKST H$\alpha$ survey of the southern Galactic plane (Parker et al. 2005), we discovered a number of previously unrecognised faint nebulae which could be associated with known Wolf-Rayet stars and have subsequently acquired spectra for some of these new nebulae. Also, given our previous comments on the difficulties in unequivocal identifications based only on available optical emission line ratios, it is possible that some of our lower confidence, new Galactic SNR candidates reported in Stupar, Parker & Filipović (2008) from our work, which currently have weak corroborating evidence as to their nature, may eventually turn out to be Wolf-Rayet star nebulae or even PNe, if a fainter Wolf-Rayet star or hot star is later uncovered in the central vicinity (see later discussion). For WR stars as least this is considered highly unlikely, since, according to the last (Version VII) Catalogue of Wolf-Rayet stars (van der Hucht 2001) and the last annexe (van der Hucht 2006) to this VII Catalogue, none of our significant number of newly uncovered nebulae have a match in this list of Wolf-Rayet stars or their nebulae. Also, it should be emphasized that 35% WR stars (Marston 1997) have an associated nebula. Our deep AAO/UKST H$\alpha$ survey has uncovered some additional nebula that could be associate with known WR stars but detecting filamentary nebulosities close to a WR does not necessarily mean they are related, especially if they are highly asymmetrically distributed. Additional kinematic information and other evidence may be required. As an example we consider two objects identified as G283.7-3.8 and G306.7+0.5 in our SNR candidate Catalogue (Stupar, Parker & Filipović 2008). G283.7-3.8 was disregarded as a WR star nebulosity due to the observed morphological form not being typical for this kind of object, e.g. the presence of long parallel filaments and the fact that the supposed exciting WR 17 star (van der Hucht 2001) is remote from the filaments in angular separation (~40 arcmin). G306.7+0.5, which, although possessing spectral characteristic and a morphological form that cannot rule out a WR star nebula origin, has been disregarded as being a WR star nebula (of the nearby WR star 53) due to the presence of an X-ray source and pulsar in the vicinity of the observed emission cloud. Consequently, both nebular structures have been classified as new optical Galactic SNRs candidates in (Stupar, Parker & Filipović 2008).

In this paper we present new spectral observations of several freshly identified nebular components in and around $\Theta$ Mus (WR 48), the second brightest known Wolf-Rayet star. The findings were made during our search for new optical Galactic SNRs (Stupar, Parker & Filipović 2008). We compare these spectral observations with previous observations found in the literature and discuss the likely origin and nature of the putative WR 48 shell nebulae which we now reject based on our new H$\alpha$ imagery. For the first time we also show the presence of apparent, almost directly opposing, collimated ionised jets close in to the WR star itself as uncovered on the quotient image (H$\alpha$ divided by SR) from the SuperCOSMOS UKST/AAO H$\alpha$ (SHS) data. The blocked down (factor of 16) quotient image for the relevant SHS survey field HA137 is shown in Fig. 1. It is clear from this that there is no obvious WR shell or ring around $\Theta$ Mus. There is extensive, scallop shaped diffuse emission mainly to the South and East that also loops back on itself at the Western extremity. It is 30 arcmin away from $\Theta$ Mus. This emission also appears connected to more diffuse nebulosity extending further to the West and indeed across much of the entire 25 sq.degree of the H-alpha survey field. We have also uncovered extensive, fine filamentary nebulosity of a completely different morphological character that does follow some of the general shape of the diffuse emission but is also found outside of this emission too and in the general vicinity of $\Theta$ Mus (marked by a red circle on Fig 1). This illustrates the complicated nature of the extant emission over a large 25 sq.degree region in an around $\Theta$ Mus.

## 2 METHOD AND OBSERVATIONS

In June 2004 we used the Double Beam Spectrograph (DBS)[1] of the Mount Stromlo and Siding Spring Observatory attached to the Australian National University (ANU) 2.3m telescope to acquire spectra of various nebular components previously thought to be associated with the Wolf-Rayet star $\Theta$ Mus. The DBS has a dichroic which feeds the blue and red arms of the spectrograph. In the blue we used a 600 lines mm$^{-1}$ grating which covered wavelengths between 3700 and 5500Å. For the red arm we used a higher resolution 1200 lines mm$^{-1}$ grating with spectral range between 6100 and 6800Å. This covers the important diagnostic nebula lines in the red while the resolution is sufficient for kinematics. The slit width was 2.5" and a resolution of 2 and 1Å was achieved in the blue and red arms respectively. Observational details are given in the Table 1. Data reduction was performed using standard spectral reduction routines from the *IRAF* package but supplemented with specially designed DS9 and *IRAF* scripts developed for more efficient extraction of 1D nebula spectra by colleague B. Miszalski[2]. For spectral flux calibration we used observations of the standard photometric stars: LTT4302 for June 12 and 14, 2004 and LTT4816 for June 13, 2004.

---

[1] http://www.mso.anu.edu.au/observing/2.3m/DBS/
[2] http://www.aao.gov.au/local/www/brent/pndr/

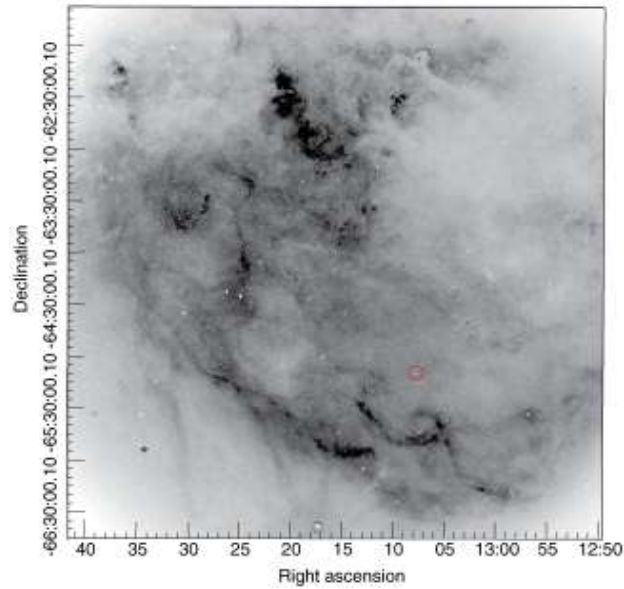

**Figure 1.** The full 25 sq.degree area of AAO/UKST H-alpha survey field HA137 as a binned 16 arcsecond per pixel low resolution quotient map of Hα divided by the matching short red (SR) broad-band image. The WR star Θ Mus (WR 48 aka HD 113904) is marked by a red circle. There is no obvious WR shell associated with WR 48. The previously identified emission shell supposedly identified with WR 48 from inferior Hα imagery in terms of sensitivity and resolution is seen to be nothing more than a more prominent arcuate emission component to the South of the star that is connected to other similar, elongated, diffuse emission structures across much of the survey field.

## 3 OBSERVING RESULTS

### 3.1 The putative Θ Muscae (WR 48) nebula: a Wolf-Rayet shell, SNR or H II region?

Narrow band, high resolution imaging from the AAO/UKST Hα survey of the southern Galactic plane (Parker et al. 2005) clearly showed the existence of various nebula structures with clearly different morphological characteristics in the vicinity of Θ Mus. One of the most prominent emission regions is a scallop-shaped nebula some 40' in E-W extent about 40 south of Θ Mus that is however, not isolated but connected to a series of similar diffuse emission structures across much of the survey field. The new Hα imaging also reveals, for the first time, the presence of a network of apparently distinct multiple fine emission filaments typically ~10' in extent extending across much of the same vicinity, sometimes overlapping sometimes not the more diffuse structures (Fig. 2). Prior to the SHS the evident nebulosity in the region around Θ Mus has generally been accepted as representing components of a WR shell nebula (for the first time reported by Heckathorn & Gull 1980). Fainter, extensive emission and finer detail of the putative WR shell have now become evident from the improved SHS imagery (see Fig. 1) which show the previously identified, supposed WR partial shell to be actually clearly connected to an extensive network of diffuse emission structures across much of the SHS survey field - see Fig. 1. There does not appear to be a distinct WR emission shell. Furthermore, careful examination of this region in the radio, particularly in the PMN survey (Condon, Griffith & Wright 1993) at 4850 MHz clearly uncovered a scallop-shaped region of radio emission in a similar form to that seen for the finer emission filaments seen in Hα light that overlaps and appears to follow the similar shaped more diffuse emission but may be unconnected or interacting with it. Fig. 3 shows this nebula at 4850 MHz. Note that the Hα filaments seen in Fig. 3 are not distinguished from the radio emission, most probably due to the low resolution of the PMN survey (~5'). However, this newly noted radio structure can also just be recognised as a very low surface brightness feature barely above the noise in the SUMSS 843 MHz (Cram, Green & Bock 1998) survey data.

For our spectroscopic follow-up we focused on several of the fine-scale newly uncovered Hα filaments (Figs. 2 and 4) as such structures are typical of optical supernova remnant detections. The observations were made as part of our spectroscopic follow-up programme of new optical SNR candidates (as example, see filaments similarity in Blair et al. 1999; Ghavamian et al. 2001. Spectra were taken at two different filament locations (see Table 2). The position and orientation of the slits is shown in Fig. 4.

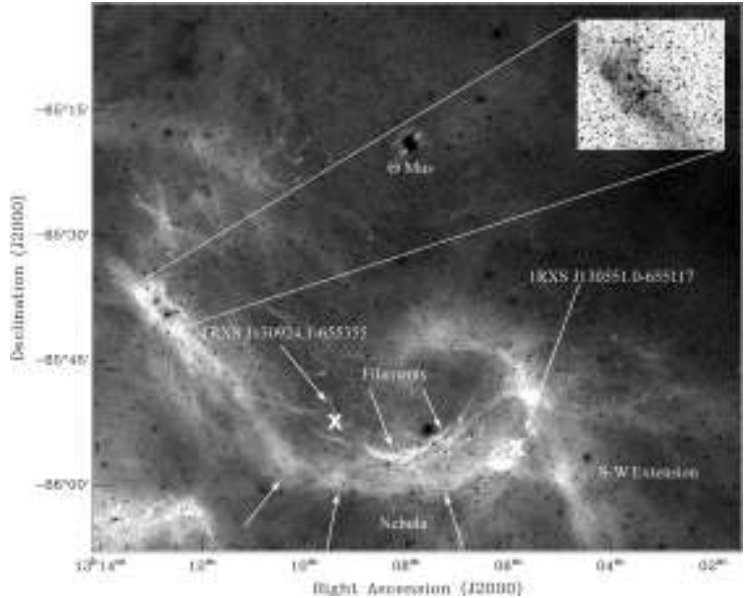

**Figure 2.** The area ~50' south the WR star Θ Mus again from the AAO/UKST Hα Survey as a binned 16 arcsecond per pixel low resolution quotient image. The fine filaments and surrounding diffuse (scallop shaped) nebula are clearly seen. The X-Ray sources 1RXS J130924.1-655355 and 1RXS J130551.0-655117 are also marked (with **X**) and lie very close to the fine Hα filaments (see later discussion). What can also be noticed on this quotient image are apparent, ionised, opposing jets close to the Θ Mus star itself (shown in greater detail in a later figure). This unique discovery is elaborated on in Section 3.3. In the right corner of the figure is an enlarged part of the image north-east from Θ Mus illustrating the fine detail in the arcuate filaments and cloud emission.

Examination of the resultant 1-D flux calibrated spectra of the two observed filaments in the vicinity of Θ Mus (see Figs. 5 and 6) showed that both spectra and resultant diagnostic emission line ratios actually fit well inside the normal criteria for optical SNRs (Fesen, Blair & Kirshner 1985) which generally distinguishes SNRs from H II regions and most PN. Many of the same optical emission lines are present in all 3 object types and although there are strong similarities the physical processes that give rise to the ionisation are different. According to Fesen, Blair & Kirshner (1985), the main factor which characterize optical spectra of SNRs, confirming a shock, is the ratio of [S II] 6717 and 6731Å/ Hα > 0.5, though others consider a ratio > 0.7 to be better (Mathewson & Clarke 1973). Such an identification can additionally be supported by the presence of [O I], [O II] and [O III] lines as well as [N II] lines at 6548 and 6584Å and the Balmer lines.

However, in addition to such line ratios being typical for optically detected SNR and not usually in PN and never in H II region emission spectra, we also have the nebulae of WR stars. Here, we have both cases: the spectral features can be very similar to the optical spectra of SNRs so that they can be very hard to distinguish (see later discussion). This is to be expected as both ejecta processes involve the creation of shocks. In SNRs the root of the shock is in the supernova explosion, while in WR star nebulae it is in the speed of stellar wind ejecta from the WR star. Besides, the observed morphological structures of WR shells can be connected with a specific evolutionary stage of the Wolf-Rayet nebula (see Marston (1995) for details). Large, slowly expanding shells which are mostly connected with the initial phase of the WR star, when the star is in the O phase, are also often seen in the IR and as H I holes. Faster, expanding shells are also present from prior to the WR phase (when the star is a red supergiant or Luminous Blue Variable) where expansion is probably connected with initial mass ejection or due to acceleration when a later WR star is formed at the centre. In a WR wind-blown shell, filamentary structures can be created when a fast WR wind overtakes the shell of surrounding ejecta. Also, on the basis of Hα and [OII] imaging morphology of currently known WR shells, they have been classified into a few basic types (Type I to Type IV) depending on the spatial displacement of the front of Hα and [O III] emission (Gruendl et al. 2000) which can be most probably applied to the shock front of SNRs for the cases when [O III] is detected.

A basic analysis of the 1-D optical emission lines shown in Fig. 2 and 3 is presented in Table 2. Due to non-photometric nights, $F(\lambda)$ instead of the extinction corrected flux $I(\lambda)$ is given. Both slit positions provide [S II]/Hα ratios of 0.86 and 0.71 (well inside that typical of SNRs). The usual Balmer and nitrogen lines are also present. The [O II] and [O III] lines are present but [O I] at 6300 and 6364Å are not seen. This is quite common when reducing low resolution spectra as the [O I] 6300 and 6364Å lines are quite prominent in the night sky spectrum which can make proper sky subtraction difficult unless there is a significant kinematic component from the source. Even if these lines are detected they often do not present the canonical 3:1 ratio demanded theoretically due to such sky-subtraction errors. The lack of [O I] at 6300 and 6364Å in both spectra can also be explained by contamination of the underlying SNR spectrum in this region with emission from a H II region (e.g. Fesen, Blair & Kirshner (1985) for the case of G166.2+2.5) or simply an interaction between the ionized gas of the scallop-shaped feature and the filaments. One more example can be found in Fesen, Blair & Kirshner (1985) where, for two slit positions for the confirmed SNR G206.9+2.3 only one spectrum showed low level [O I] emission at 6300Å.

Figs. 5 and 6 and also Table 2 show that both spectra have emission of [Ne III] at 3867Å while the spectrum from June 12, has additional lines of He+H at 3888Å and [Ne III] +H at 3970Å. These lines, thought not common, are occasionally seen in some observed SNR filaments (e.g. Fesen, Blair & Kirshner (1985) for the case of G180.0-1.7) and also in WR nebula (Esteban et al. 1990). Besides the spectra of the newly discovered fine filaments acquired for this work and the intriguing morphological structure of the filaments which are typical of an SNR, more circumstantial evidence for our proposition that we are actually dealing with a new SNR and not WR ejecta can be seen in Fig. 2 where the X-ray sources 1RXS J130924.1-655355 and 1RXS J130551.0-655117 are situated in close proximity to the fine Hα filaments (see later discussion).

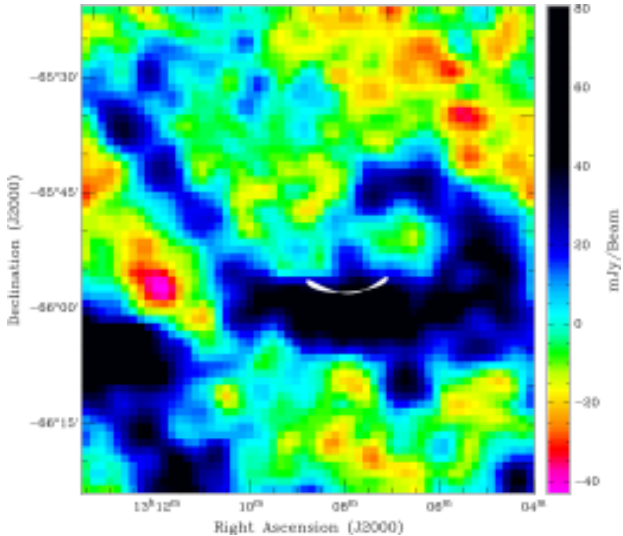

**Figure 2.** The radio image of the Θ Mus nebula region from the PMN survey data at 4.85 GHz. The inset white shape shows the position and extent of the Hα filament where the spectra were taken. Due to the low resolution of the PMN survey (~5'), the optical filament is not recognized as a separate feature in the radio. However, the overall similarity in the form of the diffuse nebula between this radio image and the optical (Hα) image shown on Fig. 2 is clear.

Our spectra of the fine optical filaments are compared with what may (or may not) be published spectra of the same filaments of Θ Mus in de Castro & Niemela (1998). Unfortunately, the exact RA/DEC values of their slit position for their spectra of the putative emission and knots from Θ Mus were not provided. However, it is clear that at least one of them[3] was taken very close (1.5' south-east) to where spectra have been taken for this work since the size of the individual filaments is typically ~10'. The basis for the spectral comparison were [S II] 6717 and 6731Å and their ratio against Hα. de Castro & Niemela (1998) used the diagnostic diagram of [S II] 6717/6731Å as a function of log (Hα/[S II]) from Sabbadin, Minello & Bianchini (1977) to show that spectra from their points (knots) are actually inside the area populated by H II regions (see Fig. 7).

We believe this is because their slits were sampling the more diffuse emission components present. For this work spectra targeted specifically on the newly discovered fine filaments in the same vicinity showed the opposite: i.e. very strong [S II] lines, with their ratio against Hα well inside the region normally occupied by SNR. Actually, Fig. 7 is an adapted diagram used by de Castro & Niemela (1998) (from Sabbadin, Minello & Bianchini (1977)) confirming strong likely SNR originated shocks in the observed filaments. The ratio of [S II] 6717/6731Å against log(Hα /[S II]) in this figure was only for comparison purposes as we use the generally accepted Fesen, Blair & Kirshner (1985) criteria for SNR classification in our work.

We also performed a check for how our spectral observations fit into the new Frew & Parker (2009) diagnostic diagrams $log$ I Hα /I [N II] versus $log$ I Hα /I [S II] where all major nebulosity groups (SNRs, PNs, Wolf Rayet shells etc.) are shown and generally well separated. However, the new plot shows that there is some overlap between classes for more extreme, evolved examples of object types. Again, however, both spectra from Figs. 5 and 6 are located well inside in the domain of evolved supernova remnants.

Chu & Treffers (1981) classified the scallop-shaped, more diffuse optical nebula prominent in previous Hα imagery and seen in better context in Fig. 2 as a radiatively excited nebula of the $R_s$ type (Chu 1981) which means that it is a shell structured H II region. On the basis of images of this area from the low-quality emission line survey of Parker, Gull & Kirshner (1979), Chu & Treffers (1981) concluded that this nebula is embedded in a more general H II region, though it is not mentioned which H II region. The only one found in the literature that covers this area is the large H II region G305.1-1.9 (size 360×135 arcmin) as listed in the comprehensive review of optical Galactic H II regions of Maršálková (1974). Note that identification of this H II region came from the low resolution Hα survey data of Dottori & Carranza (1971). Now, the whole of this region can be seen in far greater clarity from the AAO/UKST Hα data of survey field HA137 (shown in Fig.1) a which demonstrates the highly complex nature of the extensive emission structures which far exceed the reported dimensions of G305.1-1.9. Examining this survey, that of Parker, Gull & Kirshner (1979) (which was the basis for the conclusions of Chu & Treffers (1981)) and the arcminute-resolution SHASSA survey Gaustad et al. (2001) we concluded that this could have easily been identified as a single massive H II region from the point of view of these low resolution observations. However, with the availability of our high resolution SHS images (Parker et al. 2005), G305.1-1.9, where the scallop-shaped WR shell nebula is supposed to be embedded (according to Chu & Treffers (1981)), we show that this area can not be accepted as a simple H II region. The SHS data (shown in Fig. 2 and partially at large scale in Fig. 1) present this area as an extremely large complex area of emission with a concentration of different objects as might be expected for areas close to the Galactic mid-plane (at $b$ = -1.9). Some separated emission clouds could be individual H II regions. Furthermore, we failed to find in the literature any spectral observations of the 360×135 arcmin G305.1-1.9 emission cloud to accept it as a single, large H II region.

Spectroscopically, the spectra of H II regions (low excitation photo-ionised nebulae) are usually easily distinguished from SNRs, Wolf-Rayet shells and other nebulous objects including all except very low excitation PNe (due to different physical processes; see examples in Frew & Parker 2009). We conclude that accepting an area as a H II region only on the basis of low resolution optical imaging and without spectral confirmation is not satisfactory.

---

[3] de Castro & Niemela (1998) took spectra from 7 locations in the general area and it is slit position 3 in their work which appears closest to our slit position.

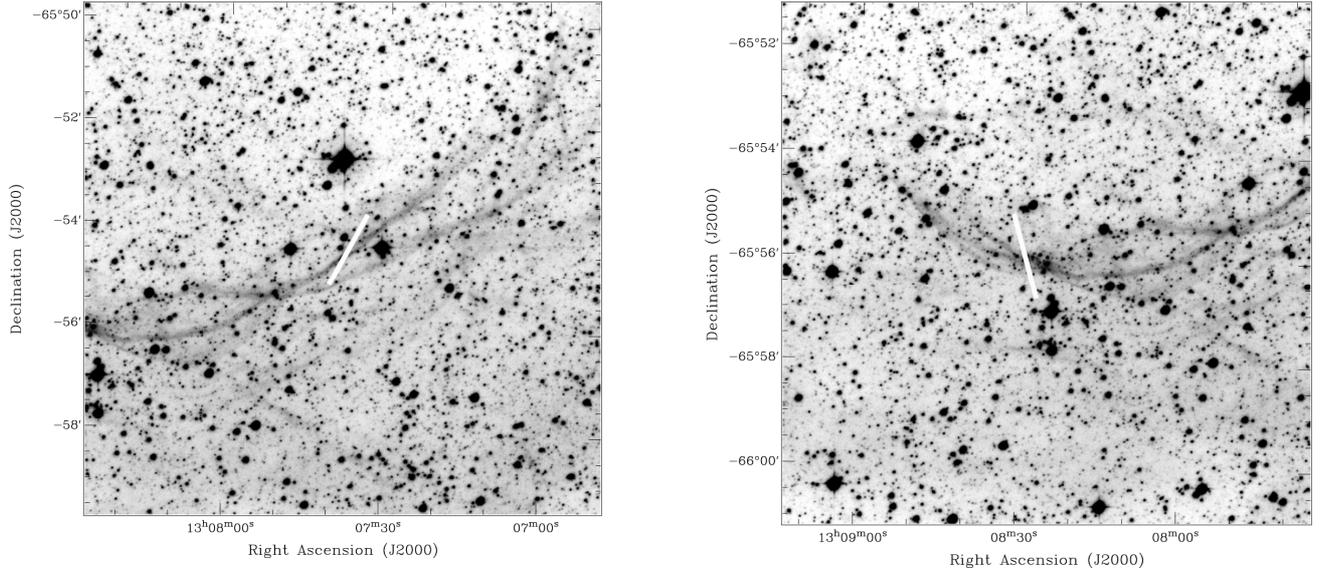

**Figure 4.** High resolution images of the area of Θ Mus from the AAO/UKST Hα survey field HA137 showing the detection of a network of fine optical filaments some 10 in extent. The white lines show the position and orientation of the 2.3-m MSSSO telescope DBS slits. Spectra were obtained on June 12, 2004 at RA(2000)=$13^h07^m38^s$ and δ = -65° 54' 37" (left image) and June 13, 2004 at RA(2000.0)=$13^h08^m26^s$ and δ = -65° 56' 06" (right image).

**Table 1**. Spectral observation log for the various optical components of nebulae in the vicinity of WR 48. All exposure were of 1200 seconds.

| Object (nebula) | Telescope | Date | Grating (lines mm$^{-1}$) | Spectral range (Å) | Slit RA | Slit Dec. |
|---|---|---|---|---|---|---|
| WR 48 | 2.3-m | 12/06/2004 | 600[a] | 3700-5500 | 13 07 38 | -65 54 37 |
|  | 2.3-m | 12/06/2004 | 1200 | 6100-6800 | 13 07 38 | -65 54 37 |
|  | 2.3-m | 13/06/2004 | 600 | 3700-5500 | 13 08 26 | -65 56 06 |
|  | 2.3-m | 13/06/2004 | 1200 | 6100-6800 | 13 08 26 | -65 56 06 |

[a] For the 600 and 1200 lines mm$^{-1}$ gratings the rms error in the dispersion solution (in Å) was between 0.05Å and 0.02Å and the relative percentage error in the flux estimate was between 10% and 17% for the 600 lines mm$^{-1}$ grating and ~20% for the 1200 lines mm$^{-1}$ grating.

We also re-evaluate the previously identified shell around the Θ Mus Wolf-Rayet star and the Chu & Treffers (1981) classification in the context of our deep, high resolution Hα imagery. Inspection of the derived quotient images, confirms the existence of the previously noted bright, extended scallop-shaped arc to the south-west of Θ Mus (see Fig. 2). Therefore, if we define a shell around a Wolf-Rayet star as comprised of "arcs of nebulosity centered on and ionized by the Wolf-Rayet star" (Chu (1981) and references therein) and accept from Chu, Treffers & Kwitter (1983) that "a WR ring nebula is an identifiable symmetric nebula around a WR star", then the observed and clearly connected south-west extension to the putative Θ Mus shell does not fit with the original morphological classification and is not what we see around other Wolf-Rayet stars (which can exhibit one or more rings). This conclusion is based on examination of the morphological structures of Wolf-Rayet shells in Miller & Chu (1993), Marston (1995) or Gruendl et al. (2000) as examples, where it is clearly seen that the shells are more or less in a circular or oval coherent form. In the low resolution, but high sensitivity SHASSA survey (Gaustad et al. 2001), even though not as clearly seen as in the SHS survey, the bright, south-west extensions from the putative WR shell can also be recognised. We conclude that the low sensitivity of previous surveys gave rise to the scallop-shaped nebula being incorrectly associated with the Wolf-Rayet star Θ Mus (WR 48). Unfortunately, de Castro & Niemela (1998) took spectra of the emission cloud south-east of the scallop-shaped nebula (emission seen in the lower left corner of Fig. 2) but not on the purported partial ring itself (see their position 5). The resultant spectra is typical for a H II region. Although, de Castro & Niemela (1998) claimed that this cloud belongs to a different component of the Θ Mus shell, it is very hard to validate their connection of this emission with the scallop-shaped nebula. On the other hand, Heckathorn, Bruhweiler & Gull (1982) claimed that, according to their imaging, in the scallop-shaped nebula [S II] (and Hβ) are not present. This is in contradiction with Fesen, Blair & Kirshner (1985) where some value of [S II] should be present (at least [S II]/Hα <0.5) to define and separate the emission of H II regions from, say, SNRs and PNs.

The nebula structures south of Θ Mus star are part of a very complex area of diffuse and filamentary emission and cannot solely be explained as resulting from ionised WR star ejecta. Indeed, cht81 concluded from kinematic and photometric distance estimates for the star itself (1.3 and 0.9 kpc respectively), from the physical diameter of the main diffuse partial shell 40 arcmin to the south (11 pc at the distance of 1 kpc), via determination of the expansion velocity of 7 km s$^{-1}$, and from the resultant lifetime of the nebula of about $10^6$ years, that this could be in fact a stellar wind nebula or supernova remnant. This conclusion follows as the lifetime of the nebula is greater then the estimated lifetime of star's Wolf-Rayet phase.

**Table 2.** Characteristic spectral lines for the first slit position in the vicinity of Θ Mus.

June 12, 2004

| Line | λ (Å) | F(λ) Hβ =100 | Line | λ (Å) | F(λ) Hα =100 |
|---|---|---|---|---|---|
| [O II] | 3727 | 1073 | [O I] | 6300 | |
| [Ne III] | 3867 | 15 | [O I] | 6364 | |
| He+H | 3888 | 19 | [N II] | 6548 | 31 |
| [Ne III] +H | 3970 | 10 | Hα | 6563 | 100 |
| Hδ | 4101 | 24 | [N II] | 6583 | 106 |
| Hγ | 4340 | 42 | [S II] | 6717 | 47 |
| Hβ | 4861 | 100 | [S II] | 6731 | 39 |
| [O III] | 4959 | 98 | | | |
| [O III] | 5007 | 271 | | | |

F(Hβ) = $2.62 \times 10^{-14}$ erg cm$^{-2}$ s$^{-1}$ Å$^{-1}$    F(Hα) = $1.40 \times 10^{-14}$ erg cm$^{-2}$ s$^{-1}$ Å$^{-1}$

[N II] / Hα = 1.37
[S II] / Hα = 0.86
[S II] (6717/6731) = 1.20

June 13, 2004

| Line | λ (Å) | F(λ) Hβ =100 | Line | λ (Å) | F(λ) Hα =100 |
|---|---|---|---|---|---|
| [O II] | 3727 | 2921 | [O I] | 6300 | |
| [Ne III] | 3867 | 168 | [O I] | 6364 | |
| Hδ | 4101 | | [N II] | 6548 | 18 |
| Hγ | 4340 | 50 | Hα | 6563 | 100 |
| Hβ | 4861 | 100 | [N II] | 6583 | 68 |
| [O III] | 4959 | 232 | [S II] | 6717 | 40 |
| [O III] | 5007 | 740 | [S II] | 6731 | 31 |

F(Hβ) = $4.69 \times 10^{-16}$ erg cm$^{-2}$ s$^{-1}$ Å$^{-1}$    F(Hα) = $1.39 \times 10^{-13}$ erg cm$^{-2}$ s$^{-1}$ Å$^{-1}$

[N II] / Hα = 0.86
[S II] / Hα = 0.71
[S II] (6717/6731) = 1.27

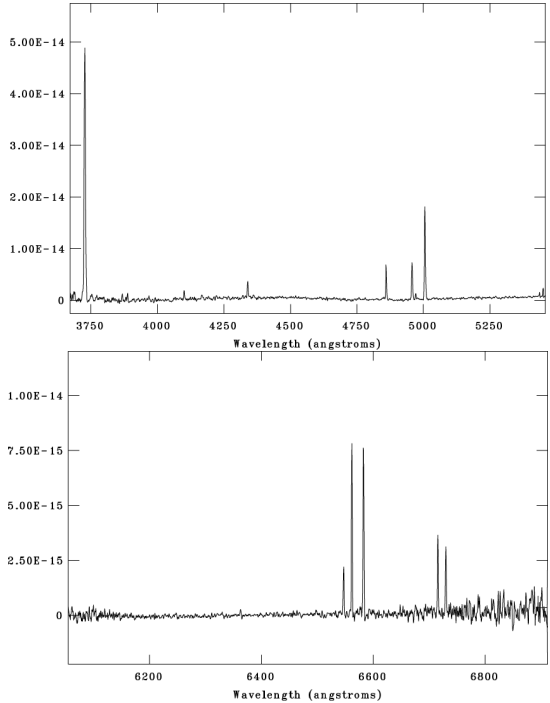

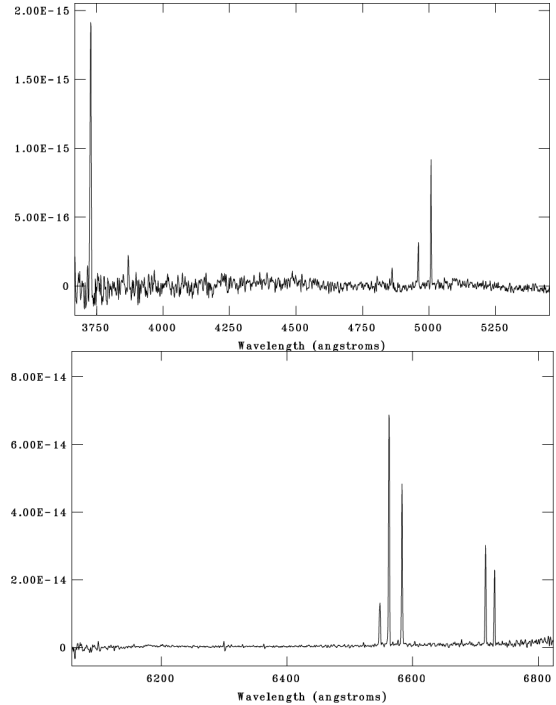

**Figure 5.** Flux calibrated 1-D blue (top image) and red (bottom image) spectra for the slit position from the June 12, 2004 observation. The high [S II] / Hα ratio of 0.9 would classify this object as being a likely SNR. In the blue spectrum, the strongest lines are of [O III] at 5007 and 4959Å and especially [O II] at 3727Å. Several Balmer lines are also recognized together with [Ne III] at 3867Å, He+H at 3888Å and [Ne III] +H at 3970Å sometimes seen in the spectra of SNRs.

**Figure 6.** Flux calibrated 1-D blue (top image) and red (bottom image) spectra for the slit position from June 13, 2004. At this slit position, [S II] / Hα =0.7 was obtained, again consistent with that from an SNR. In the blue part of the spectrum, [O III] at 5007 and 4959Å are seen together with extremely strong [O II] at 3727Å which is not expected for a H II region.

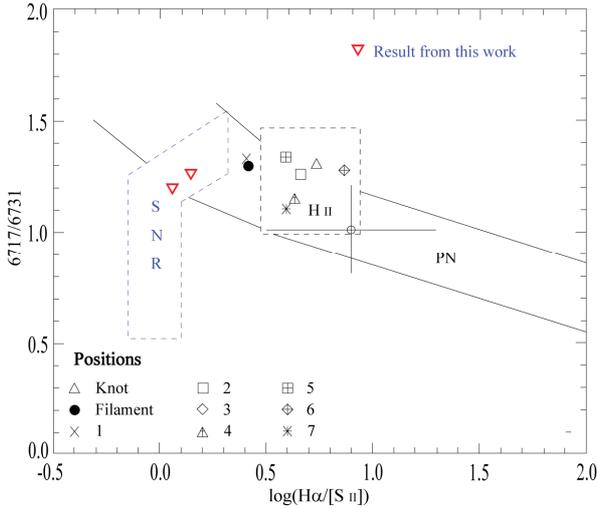

**Figure 7**. The diagnostic diagram [S II] 6717/6731 versus *log* Hα/[S II] taken from de Castro & Niemela (1998) (originally from Sabbadin, Minello & Bianchini 1977), which was used by them to classify Θ Mus as possessing an external WR nebula exhibiting H II region characteristics. However, the spectra of two newly discovered fine filaments obtained for this work fall clearly inside the area occupied by SNRs (rotated triangles). It is possible that the original spectra, lacking the high resolution quality imagery with which to select regions for study, were taken across the diffuse emission regions and not on the fine filaments now clearly visible for the first time in the high sensitivity, high resolution SHS data. Unfortunately, precise slit positions from the previously published spectra are not provided but this could easily explain the differences in spectral signatures observed.

It is also possible that the many fine filaments in the region and the adjacent much more diffuse scallop-shaped nebula (see Fig. 1), although different morphologically, are interacting. The filamentary structures, which are very reminiscent of those associated with optical SNR detections, but also seen in Dufour, & Buckalew (1999) as part of Wolf-Rayet nebula, are first found immediately to the east and west of the star but also occur further south and west in a broad sweep of filaments. These certainly seem to follow the general curvature of the much more diffuse broad nebulosity arc before then extending further west. Perhaps this is happening as a consequence of different processes of nebula in this area and the possible SNR what is implied by our spectral observations.

### 3.2 Proposed new optical components of a new Galactic SNR G304.4-3.1

The fine SHS structures in Fig. 2 and Fig. 4 seen for the first time in such detail south of Θ Mus have given us the opportunity to more clearly resolve the various morphologically distinct filaments and diffuse emission structures that appear to be present and which most probably arise from different physical processes. We believe that the existing multi-wavelength optical, radio and X-ray data now available, together with our new spectral evidence supports the notion that these fine filaments and associated spectra indicate the presence of a previously unrecognised SNR separate from the diffuse emission regions. There is no doubt that the filaments (see also Hester (1987)), seen on Fig. 4 as "ropes" (another possibility are thin "sheets" seen edge-on) are typical for the shock front of supernova remnants.

Further, the work of Marston (1996) uncovered several large IRAS 'bubbles' in the area of known Wolf-Rayet stars, including a bubble in the Θ Mus region which is clearly seen in the IRAS 60μm image presented in Fig. 8 (left plate)[4] with marked approximate position of Hα filaments from Fig. 4. Keeping in mind that the IRAS 60μm and PMN 4.85 GHz are both low resolution surveys (2 and 5 arcmin typical resolution respectively) we overlaid PMN contours over the IRAS 60μm bubble (right on Fig. 3.2) and found an excellent positional match on the south-east side of the image between the IRAS and PMN data[5]. Although the PMN signal is fragmented, a match between emissions can also be noticed on the south-west side. This identification of an IR bubble and almost the same component in the radio can also support our proposal that this is a case of a new supernova remnant. Regarding the structure of the radio emission at 4.85 GHz, shown on Fig. 8, it is in accordance with our analysis of many Galactic SNRs in the PMN survey (partially elaborated in Stupar et al. (2005)) and confirms that this morphological structure of fragmented radio signals (often very little over the level of noise) is very common between known but senile SNRs registered in the PMN. IR studies of SNRs are mostly connected with highly evolved (and large) remnants with weak radio detection expected due to low shock velocities (Saken, Fesen & Shull (1992) and references therein). One more item of support comes from the presence of two X-ray sources (Fig. 2) where 1RXS J130551.0-655117 is nearest to the center of this new proposed SNR (see Fig. 8) which we estimate to be around R.A.(2000)=$13^h 05^m 31^s$ and δ = -65° 55' 47" or at G304.4-3.1. The arcuate filaments from Fig. 4 are partially opened towards the X-ray source 1RXS J130924.1-655355. However, this source cannot easily be accepted as the centre of this newly proposed SNR due to their small extension of ~10 arcmin for these filaments compared with the overall size of G304.4-3.1 (~1.7°; see Fig. 8). If there is no connection between these X-ray sources and G304.4-3.1, some link with the pulsar J1306-6617 (Fig. 8) is possible, though the pulsar is not close to the apparent centre of the SNR due to the expected (proper/transverse) motion of the star. Furthermore, very old SNRs seen in the optical (mostly Hα), are usually fragmented inside their radio borders (if any radio emission is detected) in different forms, from filaments to emission clouds, and are probably regularly shaped by local processes within the ISM. There is clear similarity between these filaments, (east and south-east from Θ Mus, see Fig 2) and randomly distributed optical filaments in other SNRs, e.g. G279.0+1.1 (see Stupar & Parker 2009), inside ~2° radio border, and G332.5-5.6 (Stupar et al. 2007c) inside ~30' radio border.

A systematic investigation is required to get the final conclusion about the true nature of this complex of nebular structures. However, our new data do support re-classification of the extensive fine, newly detected optical filaments as resulting from an SNR. More optical spectra for additional components of the diffuse nebula and the fine Hα filaments are required. The nature of the X-ray source 1RXS J130551.0-655117 (and 1RXS J130924.1-655355) should also be examined together with the relevance of the nearby pulsar J1306-6617 and the possible confirmation of non-thermal nature of the weak radio detections. All this give additional weight to the proposal for an SNR in the region which we believe we have detected in the form of fine optical Hα filaments, confirmatory optical spectra as well detection in the IR and radio regimes.

---

[4]This bubble structure can also be noticed at the other IRAS wavelengths.
[5]Unfortunately, a check in the SUMSS 843 MHz data showed that only part of the examined area is covered in this survey.

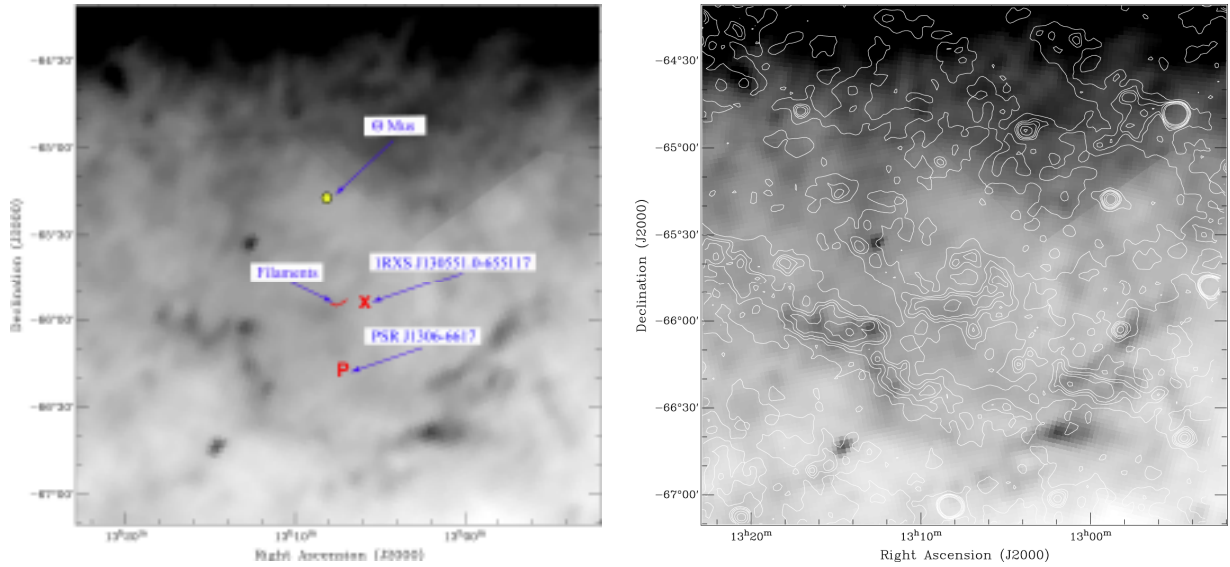

**Figure 8.** IRAS 60μm image showing a large dust hole around the Θ Mus area (left image) with the approximate position of the Hα filaments from Fig. 4 indicated. The right image is also the IRAS 60μm image but overlaid with PMN 4.85 GHz radio contours (up to 0.08 Jy beam$^{-1}$). Although the signal is fragmented, on the south-east side there is a clear match between the IRAS and PMN signals as well as partial match of radio emission on the south-west side of the IRAS bubble. One can notice the position of Θ Mus close to the edge of the shell. If we really have a Wolf-Rayet shell, Θ Mus should be, if not in the centre of the shell, then close to the centre which is not the case.

For now this proposed new remnant, G304.4-3.1, is only seen in the radio in the PMN survey at 4850 MHz. Clearly establishing non-thermal emission from this remnant is the best way of confirmation. However, in such case as this, when the remnants are old and evolved, they are usually highy fragmented in the radio and mixed in with background noise and not easily recognised. Therefore, we need a range of high sensitivity radio frequency observations to confirm non-thermal emission. It is clear from Fig. 4 that these fine filaments are completely consistent with those seen in other evolved supernova remnants (see previous discussion) but not to young SNRs, where [S II] ratio of 6717/6731Å should be ~ 0.5 and not for low density ~ 1.2 what follows from our spectral observations and what is common for evolved SNRs.

### 3.3 Possible optical jets from Θ Muscae

During examination of the complex nebulae around the Θ Mus star from the high resolution SHS images, unusual, low surface brightness emission spikes were seen emanating from the star as directly opposing, quite well collimated "jets" with narrow cone angles in approximate SE and NW directions. An enlarged 5×5 arcminute quotient image of the full resolution data centred on Θ Mus of Hα over SR is shown in Fig. 9 which illustrates this clearly. The NW jets (four relatively strong and one lower in brightness) have a radial maximal extension some 140 arcseconds from the centroid of the stellar image though there is a suspicious drop off in jet intensity at the edge of the diffraction halo for the NW emission though the SE jet extends past the main diffraction halo of the star before the intensity drops off. The extension of the three SE jets is shorter, only some 78 arcseconds with a slightly broader opening cone angle. There are also a couple of weaker but equally narrow jets at PA of ~20° in an essentially easterly and westerly direction from the star's centre. To clarify the uniqueness of these jets, we checked all other southern Wolf-Rayet stars visible in the SHS, and did not notice any similar effects. One of us (QAP) has 25 years of experience with UKST imagery and has extensively studied the SHS and its associated quotient imaging and has not encountered an equivalent feature before. There are several points to appreciate here. Firstly, the Hα emission revealed in collimated form close-in to Θ Mus is a unique feature not seen in any other of the hundreds of stars of similar magnitude across 25 sq.degrees of the survey field in question. Secondly, of all the stars in the field it is the WR star that is exhibiting the feature which would be a cruel co-incidence if it is indeed some weird photographic artifact or result of some dust speck on the photographic detector at the star's position. Thirdly, the position angle of the narrow emission cones has an intriguing connection with the modeling of what is in fact a triple star system involving Θ Mus undertaken by Hill, Moffat & St-Louis (2002). They describe a cone shaped emission region that partly wraps around the OB star. It is well known that disks, filaments and jets are common with Wolf-Rayet stars (Underhill 1994) supporting emission-line spectra but this is the first time that apparent WR star jets have been seen optically. We tried to fit these jets in the model of Hill, Moffat & St-Louis (2002) where they observed dramatic variations of C III λ5696 emission-line profiles and made geometrical models that assume that emission arises from two regions: an optically thin spherical shell around the WR star and a cone-shaped region with a position angle of $\theta$=51°, strikingly similar to the position angle values for the optical emission cone seen. They concluded that the cone-shaped region is the result of collision between the WR star ejecta/winds and an OB companion star making a binary system (Θ Mus is actually a triple system; see Sugawara, Tsuboi & Maeda 2008). Beeckmans et all. (1982), observing the C IV line in the UV, also found variations of mass loss from the system most probably due to the positions of the stellar components in the Θ Mus triple system. We need improved observations to establish the possible connection between the Θ Mus triple system and any mass loss seen in the form of jets in Fig. 9.

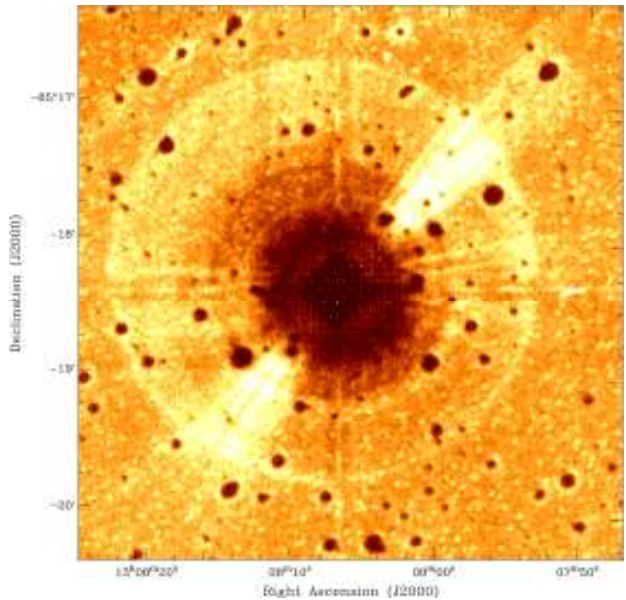

**Figure 9**. An enlarged quotient image (Hα divided by SR) of the Θ Mus star as provided from the SHS data. Several narrow jet spokes are seen to emanate from the star in almost diametrically opposing SE and NW directions. The NW jet has an apparent extension (at the angle of ~49°) of ~140 arcseconds from the edge of the photographic core of the stellar image. The extension of the jet(s) on the SE side is somewhat less at ~78 arcseconds. Faint further narrow emission rays emanating from the central star are also seen in an approximately E-W direction.

However, we stress that it is likely that the apparent angular jet extent may be a diffraction effect related to actual intense, collimated Hα emission emanating very close in to the bright star. The observed jets and their apparent angular extent are not a true physical representation of the jets that we believe may be present. This supposition is re-inforced by the visual examination of two B-grade exposures of the same SHS survey field taken weeks before on the UKST where no evidence of these jets was seen. Although the image quality of these B-grades was inferior the observed angular extent of the jets implies longevity and these jets should have been visible on the B-grades if real in a physical dimension sense. If a diffraction effect caused by intense emission then this emission could have been due to a specific outburst and does not appear to be present all the time if real. Narrow-band CCD monitoring, perhaps with an occulting disk to mask the light from the very bright star itself, is required to establish the true nature of this intriguing structure which appears unique amongst such stars. We are unable to obtain such repeat Hα imaging with the same instrumental set up on the UKST as this mode of operation is no longer supported. We present these intriguing result in the hope the community is able to shed light on this possible discovery.

## 4 CONCLUSIONS

We find that the putative Θ Mus shell previously identified in the literature is not supported by our deeper Hα imaging and related data. Careful study reveals a complex picture of extensive, overlapping and possibly interacting optical emission components from the highly diffuse to the extremely filamentary over most of the 25 sq.degree of the relevant SHS survey field HA137, much of it associated with the previously identified H II region G305.1-1.9 which itself is shown to be more extensive and incoherent. This makes it difficult to untangle the various components that are present and to determine which, if any, of the nebulae are associated with the WR star itself. Filaments and cloud emissions could originate from an SNR or the WR star or, in the case of the extensive diffuse components part of a major H II region complex. Our spectral observations confirm the presence of shocked material from the newly discovered fine emission filaments which is common only in supernova remnants though also present in some WR nebulae and evolved PN. Based on our new data we conclude that Θ Mus (WR 48) should be reclassified as a Wolf-Rayet star without a clearly identified associated WR nebula. We also propose the existence of a new, optically detected SNR in the final dissipation phase and have compiled corroborating fresh spectroscopic, morphological, radio, X-ray and infrared evidence to support this newly uncovered Galactic SNR G304.4-3.1 which is at least 45 arcminutes across.

For the first time, we have also found evidence of collimated emission cones from Θ Mus apparently in agreement with the modeling of Hill, Moffat & St-Louis (2002). These could be a cruel photographic artifact or a temporary phenomenon. Regular monitoring of the star and additional CCD narrow-band data is required to confirm the veracity of this intriguing possibility.

## ACKNOWLEDGMENTS

We are grateful referee A.P. Marston for valuable comments and suggestions that have significantly improved the paper, and to David Frew for comments given during the preparation of this paper. We are also thankful to the Mount Stromlo and Siding Spring Observatory Time Allocation Committees for enabling the spectroscopic follow-up to be obtained. We thank the WFAU of the Royal Observatory Edinburgh for the provision of the SHS data on-line and inspection of original SHS films at Plate Library of the Royal Observatory Edinburgh. We also thank Sue Tritton and Mike Read of the WFAU for the visual checking of the B-grade exposures of the Hα survey field containing Θ Mus.

## REFERENCES

Beeckmans F., Macchetto F., Grady C.A., van der Hucht K.A., 1982, in Wolf-Rayet stars: Observations, physics, evolution, IAU Proc. 99, edit. C.W.H. de Loore and A.J. Willis, Dordrecht, D. Reidel Publishing Co., p. 311

Blair W.P., Ravi S., Raymond J.C., Knox S.L., 1999, AJ, 118, 942

Cantó J., 1981, in Kahn F.D. ed., Astrophys. Space Sci. Library, Investigating the Universe. Reidel, Dordrecht, p. 95

Chu Y.-H., 1981, ApJ, 249, 195

Chu Y.-H., Treffers R.R., 1981, ApJ, 250, 615

Chu Y.-H., Treffers R.R., Kwitter K.B., 1983, ApJS, 53, 937

Condon J.J., Griffith M.R., Wright A.L., 1993, AJ, 106,, 1095

Cram L.E., Green A.J., Bock D.C.-J., 1998, 15, 64

de Castro G.C., Niemela V.S., 1998, MNRAS, 297, 1060

Dottori H., Carranza G., 1971, Ap&SS, 13, 180

Dufour R.J., Buckalew B., 1999, Proc. of the 193rd symposium of the International Astronomical Union, Eds. Karel A. van der Hucht, Gloria Koenigsberger, and Philippe R. J. Eenens. San Francisco, Calif.: Astronomical Society of the Pacific, p.350

Esteban C., Vilchez J. M., Manchado A., Edmunds M.G., 1990, A&A, 227, 515

Fesen R.A., Blair W.P., Kirshner R.P.,1985, ApJ, 292, 29

Frew D.J., 2008, PhD Thesis, Macquarie University, Sydney

Frew D.J., Parker, Q.A., PASA, in press

Gaustad J.E., McCullough P.R., Rosing W., Buren D. Van, 2001, PASP, 113, 1326

Ghavamian P., Raymond J., Smith R.C., Hartigan P., 2001, ApJ, 547, 995

Gruendl R. A., Chu Y.-H., Dunne B. C., Points S. D., 2000, AJ, 120, 2670

Heckathorn J.N., Gull T.R., 1980, BAAS, 12, 458

Heckathorn J.N., Bruhweiler F.C., Gull T.R., 1982, ApJ, 252, 30

Hester J.J., 1987, ApJ, 314, 187

Hill G.M., Moffat A.F.J., St-Louis N., 2002, MNRAS, 335,1069

Johnson H.M., Hogg D.E., 1965, ApJ, 142, 1033

Marston A.P., Yocum D.R., Garcia-Segura G., Chu Y.-H.

Marston A.P., 1995, AJ, 109, 1839

Marston A.P., 1996, AJ, 112, 2828

Marston A.P., 1997, ApJ, 475, 188

Maršálková P., 1974, ApSS, 27, 3

Mathewson D.S., Clarke J.N., 1973, ApJ, 180, 725

Miller G.J., Chu Y.-H., 1993, ApJS, 85, 137

Miszalski, B, Parker, Q.A, et al., 2008, MNRAS,

Moffat A.F.J., Seggewiss W., 1977, A&A, 54, 607

Pannuti T.G., Schlegel E.M., Lacey C.K., 2007, AJ, 133, 1361

Parker R.A.R., Gull T.R., Kirshner R.P., 1979, NASA-SP434

Parker Q.A., Phillipps S., Pierce M.J., Hartley M., Hambly N.C., Read M.A. et al., 2005, MNRAS, 362, 689

Parker Q.A., Acker A., Frew D.J. et al., 2006, MNRAS, 373, 79

Pierce M.J., Frew D.J., Parker Q.A., Köppen J., 2004, PASA, 21, 334

Sabbadin F., 1977, A&A, 54, 915

Sabbadin F., Minello S., Bianchini, A., 1977, A&A, 60, 147

Saken J.M., Fesen R.A., Shull J. M., 1992, ApJS, 81, 715

Shull J.M., Fesen R.A., Saken J.M., 1989, ApJ, 346, 860

Smith L.F., Batchelor R.A., 1970, AuJPh, 23, 203

Stupar M., Filipović M.D., Parker Q.A., White G.L., Pannuti T.G., Jones, P. A., 1995, Ap&SS, 307, 423

Stupar M., 2007, PhD Thesis, Macquarie University, Sydney

Stupar M., Parker Q.A., Filipović M.D., 2007a, MNRAS, 374, 1441

Stupar M., Parker Q.A., Filipović M.D., 2007b, AAO Newsletter, 112, 12

Stupar M., Parker Q.A., Filipović M.D., Frew D. J., Bojičić I., Aschenbach B., 2007c, MNRAS, 381, 377

Stupar M., Parker Q.A., Filipović M. D., 2008, MNRAS, 390, 1037

Stupar M., Parker Q.A., 2009, MNRAS, 391, 179

Underhill A.B., 1994, ApSS, 221, 383

Sugawara Y., Tsuboi Y., Maeda Y., 2008, A&A, 490, 259

van der Hucht K.A., 2001, NewAstr. 45, 135V

van der Hucht K.A., 2006, A&A, 458, 453